\documentclass{emulateapj}
\usepackage{natbib}

\newcommand{\kms}{\,{\rm km\,s^{-1}}}
\newcommand{\msun}{\,{\rm M_\odot}}
\newcommand{\beq}{\begin{equation}}
\newcommand{\eeq}{\end{equation}}
\def\lta{\mathrel{\rlap{\lower 3pt\hbox{$\sim$}}\raise 2.0pt\hbox{$<$}}}
\def\gta{\mathrel{\rlap{\lower 3pt\hbox{$\sim$}} \raise 2.0pt\hbox{$>$}}}
\def\simlt{\mathrel{\rlap{\lower 3pt\hbox{$\sim$}}\raise 2.0pt\hbox{$<$}}}
\def\simgt{\mathrel{\rlap{\lower 3pt\hbox{$\sim$}} \raise 2.0pt\hbox{$>$}}}
\begin{document}
\bibliographystyle{apj}

\title{Double quasars: probes of black hole scaling relationships and merger scenarios}
\author{G. Foreman\altaffilmark{1},  M. Volonteri\altaffilmark{1} \& M. Dotti\altaffilmark{1}}


\altaffiltext{1}{Department of Astronomy, University of Michigan, 500
Church Street, Ann Arbor, MI, USA}

\begin{abstract}
We analyze the available sample of double quasars, and investigate their physical properties. Our sample comprises 85 pairs, selected from the Sloan Digital Sky Survey (SDSS). We derive physical parameters for the engine and the host, and model the dynamical evolution of the pair. First, we compare different scaling relationships between massive black holes and their hosts (bulge mass, velocity dispersion, and their possible redshift dependences), and discuss their consistency. We then compute dynamical friction timescales for the double quasar systems to investigate their frequency and their agreement with the ``merger driven" scenario for quasar triggering. In optical surveys, such as the SDSS, $N_{double, qso} / N_{qso} \approx 0.1\%$. Comparing typical merging timescales to expected quasar lifetimes, the fraction of double quasars should be roughly  a factor of 10 larger than observed. Additionally, we find that, depending on the correlations between black holes and their hosts, the occurrence of double quasars could be redshift-dependent. Comparison of our models to the SDSS quasar catalog suggests that double quasars should be more common at high redshift. We compare the typical separations at which double quasars are observed to the predictions of merger simulations. We find that the distribution of physical separations peaks at $\sim 30$ kpc, with a tail at larger separations ($\sim 100-200$ kpc). The peak of the distribution is roughly consistent with the first episode of quasar activity found in equal mass mergers simulations.  The tail of the quasar pairs distribution at large separations is instead inconsistent with any quasar activity predicted by published simulations. These large separation pairs are instead consistent with unequal mass mergers where gas is dynamically perturbed during the first pericentric passage, but the gas reaches the black hole only at the next apocenter, where the pair is observed. 
\end{abstract}


\section{Introduction}
Interactions and mergers are an integral part of the process that builds up the galaxy population. Mergers are expected to be the dominant growth route of  galaxies in hierarchical scenarios linked to Cold Dark Matter cosmologies \citep{White}. Recent high resolution observations have shown evidence of such phenomena, revealing that many of the most massive galaxies at $z \sim 1-4$ are indeed interacting/merging galaxies \citep[e.g.,]{Chapman}. Many of submillimeter detected galaxies also show signs of nuclear activity \citep{Alexander2001,Alexander2003, Alexander2005}, and are found near radio galaxies, suggesting that quasar activity was common in dense environments at high redshift. Theoretical models indeed  link quasar activity to mergers  \citep[e.g.,][]{Haehnelt, Cavaliere,Kauffmann2000,Wyithe2003,VHM, DiMatteo}.  
If most galaxies host a central massive  black hole (MBH), as implied by the local observations, and nuclear activity is  triggered in mergers, double/binary quasars should also be common in a hierarchically evolving Universe.
The expected fraction of double quasars expected in a hierarchical scenario where mergers trigger nuclear activity was investigated by \cite{VHM}. If accretion were assumed to be triggered on both MBHs involved in a major merger (that is, mergers between galaxies with similar masses, up to a factor of 10 difference), the fraction of binary quasars would be about 10\%, in conflict with the existing data \citep{VHM} . A rough agreement with the data could be reached assuming that only the MBH hosted in the larger halo accretes and radiates during a major merger. In this scenario at least two subsequent major mergers are required to give origin to a double quasar system, the satellite MBH being activated in the first merger event.  In these models, the mean timescale over which MBHs accrete and radiate is of order $\sim 4\times 10^7$ years, much shorter than the typical time between two major mergers \citep{Somerville2001}. 

Only 16 pairs are known in the Large Bright Quasar Survey  \citep{Hewett} out of $10^4$ quasars, and among these 16, the confirmed physical associations are less than 10 \citep{Kochanek}. Even the most recent optical samples imply the same fraction of binaries, 0.1--0.4\%. Hennawi et al. (2006) analyze a sample of $\sim 59.000$ quasars in the Sloan survey, identifying only 26 close ($<50$ kpc) systems.  On the other hand, about 30\% of X-ray detected sub-mm galaxies at $z=2$ are in pairs (Alexander et al 2005). 

There are only a few well studied cases of ongoing mergers at sub-galactic scales in which each companion is an AGN; e.g., NGC 6240 (Komossa et al. 2003) and Arp 229 \citep{Ballo2004}. Both are advanced state merging systems, belonging to the class of ultraluminous infrared galaxies. Noticeably, none of the nuclei show signs of activity in the optical band, due to obscuration.  LBQS 0103-2753 is an example of an optically selected double quasar with separation below galactic scale, 3.5 kpc \citep{Junkkarinen}. 

Kochanek et al. (1999) analyzed the then available optical sample of double quasars in order to assess the correspondence between mergers and quasar activity. They concluded that the merger scenario \citep[as proposed by][]{Bahcall} was in good agreement with the double quasar population. They supported the merger scenario by noticing that the maximum separations between pairs are characteristic of scales on which tidal perturbations become important ($\sim 70$ kpc), and that the absence of small separation pairs is characteristic of dynamically evolving binaries. Mortlock et al. (1999) analyzes the same sample, providing a dynamical model that suggests that double quasars are associated with the very first stages of galaxy mergers. They compare the dynamical time-scale of the mergers, which results close to a Hubble time, to the expected quasar lifetime. Comparison of the time-scales implies that binary quasars are short-lived. They further suggest that the most likely explanation for the death of the binaries is that the in-flow of gas ceases, probably as the merger becomes more stable. 

We build-up on the inspirational paper by Mortlock et al. (1999), and analyze the larger sample of double quasars available today. We focus on the currently favored merger scenario, comparing the double quasar population to models related to merger simulations, and deriving black hole and galaxy properties using the most recent results on quasar-host correlations. The redshift-dependencies of the observationally-determined relations linking MBHs to their hosts is an important factor in determining the merging timescale. We discuss the implications of the claimed observed redshift-dependent
scaling relations in terms of quasar host and pair fraction evolution.  

\section{Sample selection}
For our sample, we select a total of 85 quasar pairs reported in the following papers: \citet{Hennawi} (H06), \citet{MyersB} (M07a), and \citet{MyersA} (M07b). The quasars in our sample are selected as physical pairs rather than lenses. 
The authors of our sample papers list several ways to establish whether a quasar pair is a lens or a binary. Characteristics of a lens include similar optical and radio flux ratios between the two images, measured velocity differences $\lesssim 1000 \mathrm{km\, s^{-1}}$, an observation of the lens candidate, an observation of more than two images, a measured time delay between images, and an identical set of spectra.  


H06, from which we take 69 out of our 85 quasar pairs, report quasar pairs observed using three samples from the Third Data Release \citep{Schneider} of the Sloan Digital Sky Survey, namely the SDSS Spectroscopic Quasar Sample, the SDSS Faint Photometric Quasar Sample, and the SDSS + 2QZ Quasar Sample. H06 select quasar pairs using three different methods. For pairs at separation $\Delta \theta < 3''$, the authors examine the SDSS in search of gravitationally lensed quasars and select pairs that do not meet the lens criteria discussed in \cite{Kochanek}. For pairs with $\Delta \theta > 3''$, the Authors further compare the emission of the two quasars at radio wavelengths. A pair must be a binary if it is composed by one radio bright image and one radio quiet image. Three of the reported pairs are confirmed as binaries using this method. For pairs at larger separations, they use color selection via the digital photometry of the SDSS. Follow-up observations were made to confirm the binary hypothesis. 

Finally, M07a report quasar pairs found in the Fourth Data Release of the SDSS. The authors use the color selection technique outlined in \citet{Richards} to distinguish quasars among $\sim 300,000$ SDSS objects.  M07a reports 111 quasar pairs with separations $\Delta \theta < 0.1'$. Of these 111 pairs, we select eight, requiring both images be at the same photometric redshift.  We use all pairs from Table 1b in M07b.  These quasars are confirmed binaries and are all included in M07a. 

\begin{figure}[thb]
\includegraphics[width=\columnwidth]{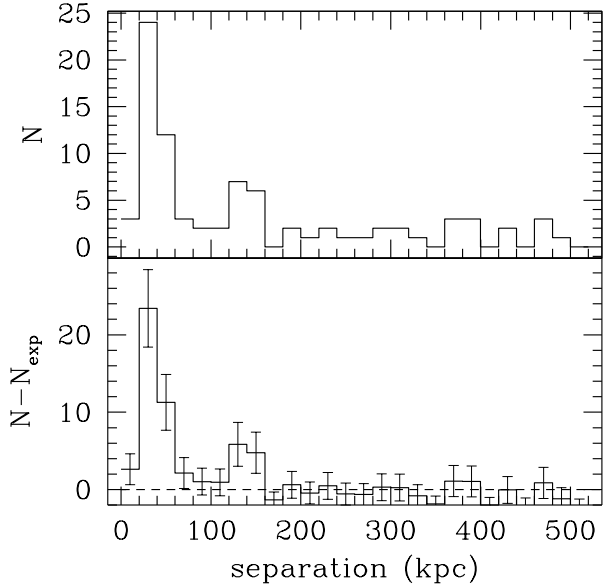}
\caption{Distribution of physical separations of double quasars. The overall distribution peaks at $\sim$ 30 kpc, with a tail extending to more than 100 kpc. The bottom panel shows the distribution when the extrapolation of the power-law quasar correlation function is subtracted (Porciani et al. 2004). Both the peak $\simeq30$ kpc, and the large separation pairs ($100-200$ pc) are above the expected number of chance superpositions.}
\label{fig:sep}
\end{figure}

Table 1 lists the data for each pair. The distribution of physical 3--d separations, assuming a concordance cosmology \citep{Spergel}, is shown in Figure \ref{fig:sep}. We do not find any trend between quasar separation and luminosity. To estimate the significance of the distribution over chance superpositions, we subtract in every bin the number of companions expected within that bin, based on the extrapolation of the quasar correlation function. For simplicity, we adopt a single power-law function $\xi(r)=(r/r_0)^{\gamma}$ \citep[where $r_0$=6.6 Mpc, $\gamma$=-1.53]{Porciani}, although our sample covers a large redshift range \citep[see also]{Hennawi}. Most of the pairs are found either at galactic separations  ($< 60$ kpc), in agreement with the results, and the arguments, of Kochanek (1999) and Mortlock (1999). The scales over which binarity is common correspond to those where tidal perturbations become important during the early phases of galaxy mergers.  A number of pairs above the expected number of chance superpositions appears also at larger scales ($\sim 150$ kpc). We will discuss these pairs further in sections 6 and 7.

\section{Sample analysis}

\subsection{Bolometric Luminosities}
Our first step consists in creating a uniform photometric sample, by deriving  $B$-band luminosities for all quasars in our sample. 

We convert the $g$- and $u$-bands to the $B$-band using a conversion appropriate to quasar spectral energy distributions: $B=g+0.17(u-g)+0.11$ \citep{Jester}. From  the apparent magnitude $B$, we derive the intrinsic luminosity $L_{B}$ in a concordance cosmology \citep{Spergel}. 
The conversion proposed by \citet{Jester} was found using synthetic photometry using quasars from the SDSS as well as from the Bright Quasar Survey. 

Bolometric luminosities $L_{bol}$ are found using the relation: 
\beq
\log L_{B} = 0.80 - 0.067 \mathcal{L} + 0.017 \mathcal{L}^2 - 0.0023 \mathcal{L}^3, 
\eeq
where $\mathcal{L}=
log(L_{bol} / \mathrm{L_{\odot}})-12$ \citep{Marconi}. Bolometric luminosities range from $6.69 \times 10^{11} \mathrm{L_{\odot}}$ to $5.94 \times 10^{14} \mathrm{L_{\odot}}$. The average bolometric correction found from the above equation is $L_{bol}/L_{B}=5.6$.

\subsection{Black Hole Masses}
For a sample of 41 quasars belonging to the Sloan Data Release 3 we can use MBH masses derived by \cite{Vestergaard2008} by adopting the scaling relations based on the broad emission line widths and nuclear continuum luminosities \citep[e.g.][]{Vestergaard2002,Warner,VestergaardB,Peterson}.  For the remaining quasars, we derive black hole masses by applying a constant Eddington ratio, $f_{Edd}$.  We compute $f_{Edd}$ on the sample compiled  by \cite{Woo}, selecting sources with luminosities in the same range as our quasars. We find a median Eddington ratio $f_{Edd}$ of 0.29 with a 0.3 dex scatter.   \cite{Kollmeier} find $f_{Edd} = 0.25$ with a 0.3 dex scatter on a sample of X-ray and mid-infrared (24$\mu$m) AGNs from the AGN and Galaxy Evolution Survey where black hole masses are measured using H$\beta$, MgII, and CIV line emission along with continuum luminosities. This result is consistent with our Eddington ratio of 0.29. Using $L_{bol} / \mathrm{L_{\odot}}=3.27 \times 10^{4}f_{Edd}(M_{BH} / \mathrm{M_{\odot}})$, black hole masses range from $7.05\pm^{13.4}_{4.76} \times 10^{7} \mathrm{M_{\odot}}$ to $6.26\pm^{12.0}_{4.31} \times 10^{10} \mathrm{M_{\odot}}$. 

As a further check on our black hole masses, we apply also the relation suggested by \cite{Peng}, based on the blue luminosity of the quasar: 
\beq
\log M_{BH,8} = 7.88 + 0.79 \times \log\left(\frac{\lambda L_{\lambda,5100}}{10^{44} \, {\rm erg\, s^{-1}}}\right),
\eeq
where $M_{BH,8}$ is the black hole mass in units of $10^8\msun$. We convert $L_{bol}$ to $\lambda L_{\lambda,5100}$ using $L_{bol} \approx 9\lambda L_{\lambda,5100}$ \citep{Peterson, Kaspi}. The black hole mass range is $1.7 \pm 1.0 \times 10^{8} \mathrm{M_{\odot}}$ to $3.7 \pm 3.4 \times 10^{10} \mathrm{M_{\odot}}$. These black hole masses are consistent within error with those found from the Eddington ratio. Finally, as a final check, for the 41 black holes with masses determined by \cite{Vestergaard2008}, the masses we derive stepping through bolometric correction and Eddington ratio are within a factor of two from the more reliable masses estimated by \cite{Vestergaard2008}. Whenever an MBH mass is available from \cite{Vestergaard2008}, we use the spectroscopically determined MBH value for our analysis.

\subsection{Host Galaxy Properties}
Since our goal is to investigate the dynamical properties of quasar pairs, we need to embed quasars and their black hole engines in appropriate hosts, providing the potential for the dynamical decay. Since measuring quasar host properties rather than with Hubble Space Telescope is daunting, we adopt here an indirect method, in which we {\it assume} that black hole masses scale with the properties of their hosts \citep{Ferrarese, Tremaine,Treu,McLure,Peng,HopkinsB}. Albeit indirect, our method allows us to consider how different correlations, and their possible redshift evolution, affect the dynamical evolution of quasar pairs. 

Using black hole masses, $M_{BH}$, we then proceed to calculate several properties of the galaxies that host the quasars in our sample, namely stellar velocity dispersion $\sigma$, stellar bulge mass $M_{bulge}$, dark matter halo mass $M_{halo}$, and dark matter halo radius $R_{halo}$. 
Our starting point for the analysis will be in one case the $M_{BH}$-$\sigma$ relation from \citet[case A]{Tremaine}, and in a second instance the $M_{BH}$-$M_{bulge}$ relation from \citet[case A]{Haring}, which does not include redshift evolution. We also test the redshift evolution of the correlation between $M_{BH}$-$\sigma$ suggested by \cite{Treu}, the $M_{BH}$-$M_{bulge}$ relation from \citet{McLure}; cases C and D. We will comment on this test in section 4.

\subsubsection{Case A: the velocity dispersion}
Our first endeavor at estimating the properties of the host stems from the correlation between black hole mass and velocity dispersion of the host bulge \citep{Ferrarese,Tremaine, Woo2008}. In particular, we adopt the correlation proposed by \citet{Tremaine}, for the more simplified definition of velocity dispersion, and the relation proposed by \cite{Woo2008} to test evolutionary effects (see Section 4).

From black hole masses we compute stellar velocity dispersions via:
\beq
\log M_{BH,8} = 0.13 +4.02\log \sigma_{200}.
\eeq
From this relation, we find stellar velocity dispersions to range from 160$\mathrm{km\, s^{-1}}$ to 540$\mathrm{km\, s^{-1}}$. 

Since one of our goals is to compare the sample of quasar pairs to the merger scenario, we relate the bulge mass to a velocity dispersion by assuming the ``fundamental plane of black holes" proposed by  \cite{HopkinsB},  where the authors use galactic merger simulations to relate MBH masses, host bulge mass and velocity dispersion. The plane defined in \cite{HopkinsB} is claimed to be redshift-independent, if quasar progenitors at each redshift have a wide range of gas fractions. Once we derive velocity dispersions from the MBH masses using the relationship from \cite{Tremaine} we use: 
\beq
\log M_{BH,8}  = 0.11 + 0.54\log M_{bulge,11}  + 1.82\log \sigma_{200}
\eeq
to find the bulge masses of the host galaxies in our sample, where $\sigma_{200}$ is the velocity dispersion in units of 200 $\kms$ and $M_{bulge,11}$ is the bulge mass in units of $10^{11}\msun$.

 From this relation we find $M_{bulge}$ to range from $3.2 \times 10^{10}\mathrm{M_{\odot}}$ to $1.2 \times 10^{13}\mathrm{M_{\odot}}$.  Assuming that the stellar velocity dispersion is a proxy for the virial velocity \citep{Ferrarese2002}, we can find the properties of the host dark matter halos. From the virial theorem and the spherical collapse model, we derive dark matter halo masses that range from $4.5 \times 10^{12}\mathrm{M_{\odot}}$ to $1.9 \times 10^{13}\mathrm{M_{\odot}}$, and the radii of the dark matter halo are in the range 190 kpc to 660 kpc. The masses of the host halos are in agreement with, although slightly higher than, the typical masses of quasar hosts derived from clustering analysis \citep[see Fig. 2]{MyersB,Coil}.

\begin{figure}[thb]
\includegraphics[width=\columnwidth]{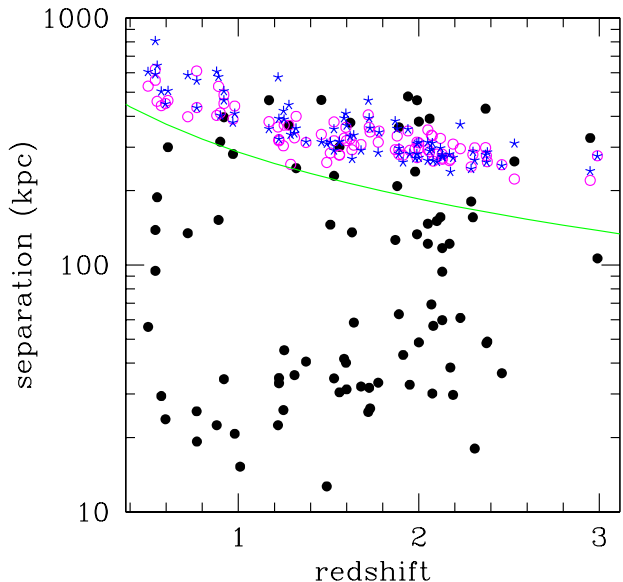}
\caption{Physical separations of double quasars as a function of redshift (solid circles). The sizes of the halos that we derive for our sample are shown as well.  {\it Open circles:} case A, velocity despersion. {\it Asteriscs:} case B, bulge masses. For comparison we show by a solid curve the size of a halo with mass $5\times10^{12}\msun$, the typical quasar host halo found in clustering analysis \citep{MyersB,Coil}. }
\label{fig:zvsD}
\end{figure}

\subsubsection{Case B: the bulge mass}
We can derive the mass of the host bulge by applying two observational, empirical relations, one that was found for local quiescent MBHs \citep{MarconiHunt,Haring} and one that was found for high-redshift radio-loud quasars \cite{McLure}. Our case B adopts the relationship proposed by \cite{Haring}:
\beq
\log M_{BH,8}=0.2+1.12\log M_{bulge,11}.
\eeq
Bulge masses are in the range $2.9 \times 10^{10} \mathrm{M_{\odot}}$ to $6.2 \times 10^{12} \mathrm{M_{\odot}}$. We will discuss the effect of adopting the relationship proposed by \cite{McLure} in the next section. 

We now reverse the method used for case A and we apply Equation 2 to compute velocity dispersions starting from the stellar bulge masses.  Velocity dispersions range from 170$\mathrm{km\, s^{-1}}$ to 840$\mathrm{km\, s^{-1}}$. We can compare the velocity dispersion that we derive ``directly" from the black hole properties, e.g., using \cite{Tremaine}, to the velocity dispersion that we derive ``indirectly" by applying in order the relationship by \citet{Haring} to find the bulge mass and then the Black Hole Fundamental Plane to find the velocity dispersion, from black hole and bulge mass. The relationship reads  $\log M_{BH,8} = 0.02 +3.50\log \sigma_{200}$, leading to a steeper dependence of $\sigma$ on BH mass. The ``indirect" method therefore results in a larger velocity dispersion at fixed BH mass for $ M_{BH}>2\times10^7$ . From the stellar velocity dispersion we once again find the dark matter halo mass, ranging  from $4.8 \times 10^{12} \mathrm{M_{\odot}}$ to $3.0 \times 10^{13} \mathrm{M_{\odot}}$. The radii of the dark matter halo range from 180 kpc to 840 kpc.


\section{Redshift evolution of MBH-host correlations}
In the previous models we have assumed that the correlations between black holes and their host are redshift-independent. This is indeed claimed, theoretically, for the black hole fundamental plane \citep{HopkinsA}. However, the two correlations between $M_{BH}$ and $M_{bulge}$, and $M_{BH}$ and $\sigma$ might evolve with the cosmic time \citep{McLure,Peng,Treu,Woo2006}.

\subsection{Case C: redshift evolution of $M_{BH}-M_{bulge}$}
\cite{McLure} use the 3C RR sample of radio-loud quasars to investigate the evolution of the correlation between black hole and bulge masses in the redshift range $0 < z < 2$. Stellar bulge masses are computed by the authors for a sample of radio-loud quasars from the 3C RR sample from $K$-band magnitude measurements, while black hole masses are found from H$\beta$, MgII, or CIV line widths. $M_{bulge}$-$z$ and $M_{BH}$-$z$ relations are found independently, and the two are combined to form the relation we use. Below $z=1$ the 3C RR sample is consistent with a black hole weighing 0.002 times the mass of the bulge. The ratio between black hole and bulge mass is found to evolve with redshift as $M_{BH}/M_{bulge}\sim (1 +z)^2$. We define a case C, where from black hole masses we compute stellar bulge masses via 
\beq
\log(M_{BH} / M_{bulge}) = 2.07\log(1 + z) - 3.09
\eeq
\citep{McLure}, and subsequently calculate velocity dispersions using Equation 4.

\begin{figure}[t]
\includegraphics[width=\columnwidth]{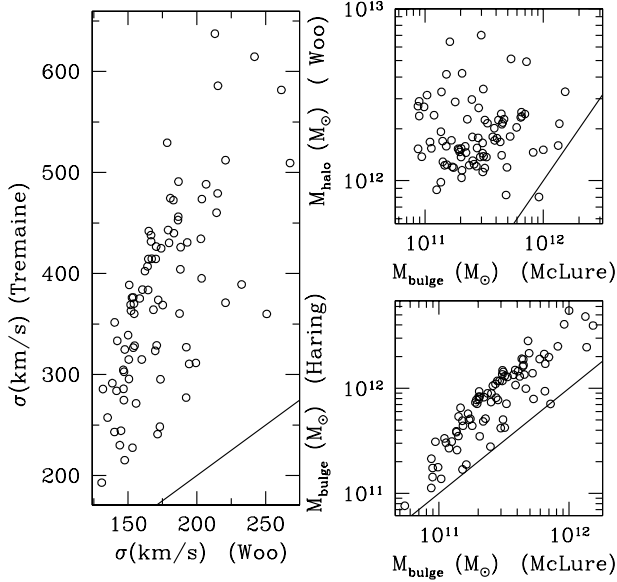}
\caption{Comparison between galaxy properties derived directly from black hole masses, using the relationships found for $z>0$ black holes (Woo et al. 2008; McLure et al. 2006). {\it Left panel:} velocity dispersions.  {\it Right bottom panel:} bulge masses. {\it Right top panel:} bulge mass vs halo mass, if we assume that the velocity dispersion can be used as a tracer of the total potential (e.g., Ferrarese 2002).}
\label{fig:comp3}
\end{figure}

\subsection{Case D: redshift evolution of both $M_{BH}-M_{bulge}$ and $M_{BH}-\sigma$}
\cite{Treu} and \cite{Woo2006} estimate black hole masses for a sample of AGN at $z=0.36$ from the H$_\beta$ line width and the optical luminosity at 5100 ${\rm \AA}$, based on the empirically calibrated photoionization method \citep[e.g.,][] {Kaspi}. They also directly measure velocity dispersions from stellar absorption lines in the host spectra. \cite{Woo2006} find that, at a fixed velocity dispersion, the mass of the MBH increases with redshift. In other words, MBHs are hosted by galaxies with smaller velocity dispersion at earlier cosmic times. At redshift $z$, 
\beq
\log M_{BH}(z)-\log M_{BH}(0)=1.66z+0.04. 
\eeq
\cite{Treu2007} further find that, for the same galaxies in the sample, MBHs are hosted by galaxies with smaller bulges, with a redshift dependence similar to what \cite{McLure} find (at least for $z\ll1$). 
The redshift scaling proposed by \cite{Woo2006} has been derived on a sample at a very specific redshift, much lower than the redshift of our quasars (see Table 1 and Figure 2). The extrapolation of their relationship at much higher redshifts leads to some ``peculiar" cases, such as our  highest redshift quasar, apparently powered by a $1.4\times 10^{10}\msun$ MBH, strikes us as being hosted by a bulge with velocity dispersion less than $50\kms$. Furthermore, in this case if we were to assume that the velocity dispersion can be used as a tracer of the dark matter potential the mass of the halo would be roughly equivalent to the mass of the bulge. 

A more recent analysis by \cite{Woo2008} modifies the redshift scaling as: 
\beq
\log M_{BH}(z)-\log M_{BH}(0)=3.1\log (1+z) + 0.05. 
\eeq
This scaling relation improves somewhat on the ``peculiar" cases mentioned above (now the host of the $1.4\times 10^{10}\msun$ MBH has a velocity dispersion of $\simeq$210 km s$^{-1}$), and the galaxy properties are now illustrated in Figure~\ref{fig:comp3}.  Still, the redshift dependence leads to very different properties for the hosts, which reflect on the dynamical evolution, when we compute  the merging timescales.  Figure~\ref{fig:comp3} illustrates these differences. If the relationships between black hole masses and their hosts are redshift-dependent, typically MBHs are hosted in smaller galaxies (both smaller velocity dispersions, and smaller bulges). In most cases, therefore, the separation between the quasars in a pair are larger than the size of their host halos, leading to the conclusion that the merger of the systems is still in its first phases, and has a long way until coalescence. 

Our case D couples the relationship by Woo et al. (2008) and \cite{McLure} to find joint velocity dispersions and bulge masses.

\section{Dynamical Friction Timescales}
From the host galaxy properties, we can compute dynamical friction timescales $t_{DF}$, i.e. the time needed for the host galaxies to merge, using the Chandrasekhar formula\footnote{Detailed calculations of the merging timescales will depend on the 
galaxy potentials and on orbital parameters of the encounters, e.g. \cite{Taffoni,Boylan}. Given the uncertainties on 
which is the most appropriate choice for the potential wells of galaxies, and the fact  that there is at most 
a factor of a few diskrepancy between the dynamical friction timescale and the merging timescale 
in the case of major mergers, we prefer a simple, clear, physically motivated choice such as the 
Chandrasekhar formula.}
\citep[M99, see also][]{Binney1987}
\beq
t_{DF} \simeq 4 \times 10^{9} (r_{max} / 20\mathrm{kpc})^{2} (\sigma_{200})/ (10^{9}\mathrm{M_{\odot}} / M)\mathrm{yr}. 
\label{eq:tdf}
\eeq
This equation is derived by integrating the equation for the time spent at a separation $r$, $|dt / dr| \simeq 3.3 (\sigma / GM \ln(\Lambda))r$, where $\ln(\Lambda) \simeq \ln(2r_{*} \sigma^{2} / GM)$ and $r_{*} \simeq 10 \mathrm{kpc}$ \citep{Lacey}. Here $\sigma$ is the velocity dispersion of the largest host galaxy, and $M$ is the mass of the infalling system. 
To find $t_{DF}$ for the galactic mergers of our sample, we use the separation of the quasars as $r_{max}$ and the largest velocity dispersion in the pair as $\sigma$. $M$ has to be related to properties of the smaller system (the ``satellite"). 
For quasars with separations larger than the size of their host halos, we assume that the merger of the systems is still in its first phases, and we expect the total time to coalescence and formation of the new galaxy to be determined by the global potential wells, hence we choose $M=M_{halo}$ in equation \ref{eq:tdf}. When instead the two quasars appear to be already ``inside" the main host halo, we use $M=M_{bulge}$, assuming that the black holes are still surrounded by their stellar envelopes \citep{Yu}. Figure \ref{fig:zvsD} compares the separations between pairs to the typical halo sizes for cases A, B. Only about 20\% of mergers appear to be still ``outside" the respective halos of galaxies.

\begin{figure}[t]
\includegraphics[width=\columnwidth]{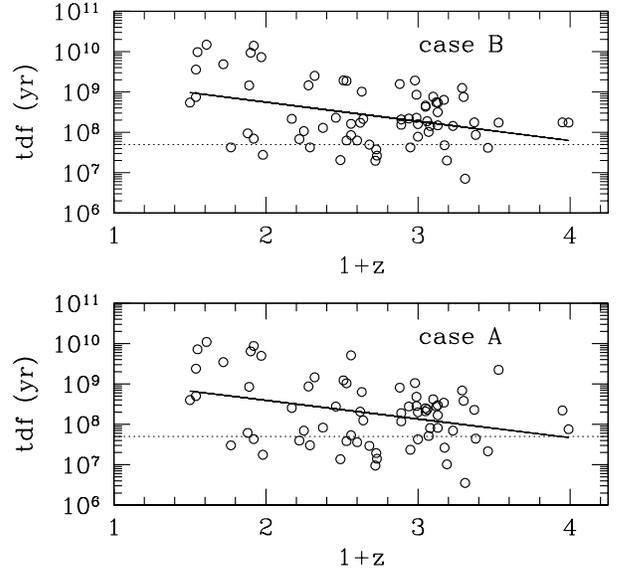}
\caption{Dynamical friction timescale for quasar pairs, in case A (bottom) and case B (top). Solid curves are linear fits. The dotted horizontal line marks the Salpeter time, $t_{\rm Salp}\sim 4.5\times 10^7 {\rm yr} (\epsilon/0.1)$.  }
\label{fig:tdfA}
\end{figure}

\begin{figure}[t]
\includegraphics[width=\columnwidth]{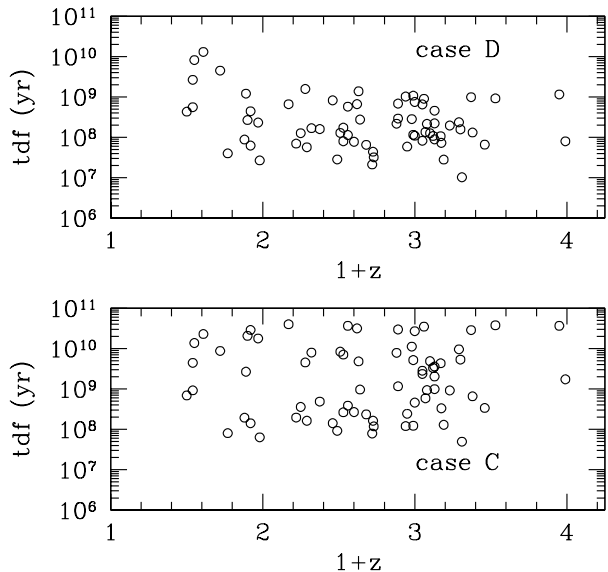}
\caption{Dynamical friction timescale for quasar pairs, in case C (bottom) and case D (top).}
\label{fig:tdfB}
\end{figure}

Our treatment is very simplified, as it does not include the possibility that we are detecting the quasars at pericenter or apocenter on eccentric orbits \citep{Khochfar}, however, it is very unlikely to detect a merger at pericentric passage, where the system spends the least amount of time, and our results are not strongly biased if the systems are near apocenter.  The timescales we find for galactic mergers are in rough agreement with simulations when we apply the appropriate radii and masses from our sample. For both cases A and B we find that the timescales range from $10^6$ yr to $10^{10}$ yr, compared to $\sim 10^{9} \mathrm{yr}$ from \citet{HopkinsA}.  The agreement is best at low redshifts, with merger timescales decreasing with increasing redshift both in case A and B.  We show the merger timescales of the quasar hosts as a function of redshift in Figure \ref{fig:tdfA}.  In our cases C and D the merger timescales vary extremely slowly with cosmic time (Figure~\ref{fig:tdfB}). The typical merging timescales remain of the order of $10^8-10^{10}$ years.

\section{Results}
We compare the merger timescales to the lifetime expectancy of quasars in order to understand the link between mergers and quasar fueling:  $N_{double, qso} / N_{qso} \simeq t_{qso}/t_{DF}$ (cfr. Martini 2004). A numerical investigation of the quasar lifetime in galaxy mergers has been presented in \cite{HopkinsA}. They suggest that the time spent at a given luminosity is ``luminosity dependent". The quasars in our sample have $L_B>10^{11}L_\odot$, which leads to $t_{qso} \simlt 10^7 \mathrm{yr}$. However, luminosities larger than $L_B>10^{11}L_\odot$ are predicted to occur only after the systems have coalesced, in the merger remnant, where the quasar separation is well below the SDSS resolution limit. Since this set of simulations is not representative of the double quasar population that we are investigating, for sake of simplicity we adopt as ``standard" value associated to quasar lifetimes the Salpeter time, $t_{\rm Salp}=\epsilon c \sigma_T/(4\pi G m_p)\sim 4.5\times 10^7 {\rm yr} (\epsilon/0.1)$, where $\epsilon$ is the radiative efficiency. If we compare $t_{qso} \sim 4.5\times 10^7 \mathrm{yr}$ to the average merging time scale, we find that $N_{double, qso} / N_{qso} \sim 4-5\%$ for case A ($\langle t_{DF} \rangle= 9.2 \times 10^{8} \mathrm{yr}$), and similarly for case B ($\langle t_{DF} \rangle = 1.3\times 10^{9} \mathrm{yr}$).


Figure \ref{fig:tdfA} also suggests that, depending on what the intimate link between black holes and their host really is, we should see a varying fraction of double quasars with redshift.  We can compute the average merging timescale in 3 redshift bins, and repeat the same analysis. In case A, for quasars at $z<1$ we find $\langle t_{DF} \rangle= (4.1 \pm 6.2)\times 10^{9} \mathrm{yr}$, yielding a binary fraction $N_{double, qso} / N_{qso} \lta 1\%$.  For quasars at $1<z<2$ the fraction increases, as $\langle t_{DF} \rangle= (3.5 \pm 5.6)\times 10^{8} \mathrm{yr}$, yielding a binary fraction $N_{double, qso} / N_{qso} \sim 3\%$. In our last redshift bin, $2<z<3$, the average merging timescale is similar to the previous bin, but the dispersion decreases, as more quasars have similar, short, merging timescales: $\langle t_{DF} \rangle= (3.2 \pm 3.9)\times 10^{8} \mathrm{yr}$, so  $N_{double, qso} / N_{qso} \sim 3\%$. Results are similar for case B. In our cases C and D the merger timescales are roughly redshift-independent, leading to the expectation of an approximatively constant binary fraction

\begin{figure}[t]
\includegraphics[width=\columnwidth]{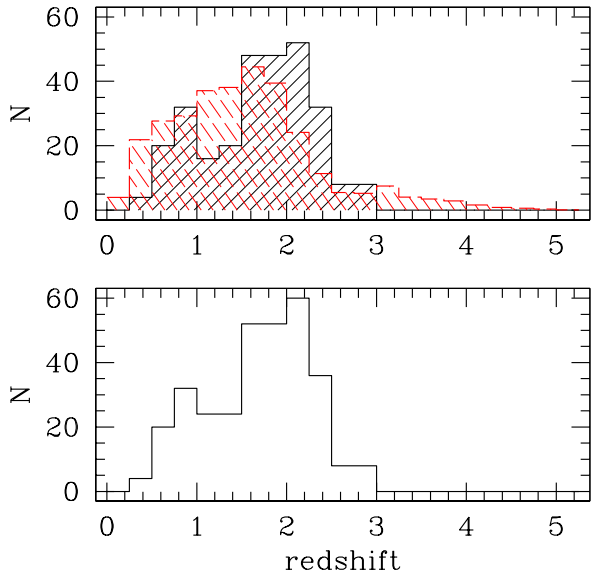}
\caption{Redshift distribution of double quasars in our sample. {\it Bottom panel:} all quasars in the sample. {\it Top panel:} double quasars from Hennawi et al. (2006) only (solid histogram). For comparison we show the redshift distribution of all quasars in the Fifth Data Release of the Sloan Digital Sky Survey quasar catalog (Schneider et al. 2007, dashed histogram), renormalized by a factor of $10^{-3}$.}
\label{fig:tdfvsz}
\end{figure}

We postpone a detailed analysis of the redshift dependence of quasar pairs to a forthcoming paper, for now we show in Figure \ref{fig:tdfvsz} some tentative evidence that the pair fraction might indeed be redshift dependent.  Here we can calculate the binary fraction in the same redshift bins where we have compared $t_{qso}$ to the merging time scale. We find $N_{double, qso} / N_{qso} \simlt 0.1\%$ at  $z<1$, and also at  $1<z<2$, while  $N_{double, qso} / N_{qso} \sim 0.2\%$ at  $2<z<3$.

Assuming that only one physical process triggers quasars, we can compare the overall redshift distribution of quasars in a homogenous statistical sample to the distribution of pairs only. We select only double quasars from Hennawi et al. (2006) and compare their redshift distribution to that of all quasars in the Fifth Data Release of the Sloan Digital Sky Survey quasar catalog \citep{Schneider}. Figure \ref{fig:tdfvsz} shows a hint of increasing pair occurrence at $z>1.5$.

Finally, we compare the typical separations at which binary quasars are observed  (Figure~\ref{fig:sep}) to the merger phase and separations that should correspond to the quasar phase in merger simulations. We find that the distribution of physical separations peaks at $\sim 30$ kpc, with a tail at larger separations ($\sim 100-200$ kpc). The peak of the distribution is roughly consistent with the first, earliest (low-luminosity) episode of quasar activity found in equal mass mergers simulations  \citep[see Figure 2 in][]{HopkinsC}, although the numerically predicted luminosities in this phase of the merger are smaller than the typical double quasar luminosity in our sample.  The tail of the quasar pairs distribution at large separations is instead inconsistent with any quasar activity predicted by published simulation results. 

Nevertheless, the mass ratio of the black holes in the pairs (that we can use as a proxy for the typical mass ratios of the merging galaxies) does not match the 1:1 ratio that was typically selected in galaxy merger simulations. The average mass ratio is $\langle 2.4\pm1.3\rangle$, which corresponds to a regrettably unexplored range. When we consider a smaller secondary galaxy, we expect a smaller deceleration due to dynamical friction, and consequently a larger galaxy separation at the first apocenter. \cite{Colpi2007} indeed show that in a 1:4 galaxy merger \citep{Kazantzidis} the first apocenter occurs at approximately 180 kpc. 

The presence of the large separations tail suggests that a significant fraction of double quasars is seen at the first apocentric passage, where the two galaxies spend a much longer time with respect to the pericenter. This scenario would follow naturally if gas is dynamically perturbed during the first pericentric passage, forming soon after a gaseous circumnuclear disk \citep[with a size of a few hundred parsecs, e.g.][]{Kazantzidis}, but a delay occurs before the gas in the disk looses enough angular momentum to be driven at subparsec scales where it can finally be accreted by the MBH. This final stage likely occurs during the first apocenter, where the quasars are observed, causing the tail at $\sim 100-200$ kpc.











\section{Conclusions}
If most galaxies host a MBH \citep[e.g.,][]{Richstone2004} and galaxy mergers trigger quasar activity (Haehnelt \& Rees 1993; Wyithe \& Loeb 2003;  Kauffmann \& Haehnelt  2000; Cavaliere \& Vittorini 2000; Di Matteo et al. 2005) then the observational sample of binary quasars can be used as a probe of quasar lifetimes. Let's start with Occam's razor and assume that every galaxy hosts a MBH and quasar activity is triggered by galaxy mergers. If  the lifetime of quasars equals the merger timescale, $t_{qso}=t_{DF}$, the probability of observing a double quasar is in principle 100\%, if we do not consider additional factors, such as obscuration. If $t_{qso}\ll t_{DF}$, then the probability  much smaller than 100\%. Additionally, if there is a delay in the triggering of the two quasars, then one might have ceased its activity before the other started, for instance. Face value, our results require that the lifetime of quasars is extremely short, about an order of magnitude shorter than the Salpeter time, or the lifetime expected from numerical simulations.

In a merger driven scenario for quasar activity, with MBHs in most galaxies, where naively all sources are expected to be in binaries, the 0.1\% binary fraction can be explained in different ways:
\begin{itemize}
\item Quasar pairs are obscured much more than expected from simulations, at Compton thick levels ($N_H>10^{24}{\rm cm^{-2}}$), otherwise pairs would be easily detected in X-rays.
\item Activity is triggered on only one of the 2 black holes, or on the MBH(s) in the merger remnant. Even if there is a MBH binary in the remnant (Hopkins et al. 2006), we should expect the binary to be hard, or at least the MBHs to be bound by the time the galaxy merger remnant has relaxed enough to be considered a "normal galaxy". A pc-scale binary would not be resolved in the SLOAN or LBQS surveys. However, this hypothesis does not explain pairs occuring  at ten-kpc or hundred-kpc scales.
\item There is a time delay between the onset of activity in the 2 merging systems (e.g. the gas infall timescale depends on the dynamical time of each host, rather than the dynamical time of the merging system, see \citet{Colpi2007}). If  the two quasars light up at different, random times, then  $N_{double, qso} / N_{qso} \simeq (t_{qso}/t_{DF})^2\simeq0.1$\%, consistent with the observational result.  If the time delay is longer that the quasar timescale, then binarity is avoided completely.
\end{itemize}

The average mass ratio of the quasar pairs we analyzed is $\simeq2$, while, since the mass function decreases steeply at large galaxy masses it is more probable that a very massive galaxy\footnote{As all the MBHs in our sample have masses above $10^8\msun$, they are hosted in large galaxies sitting at the steep end of the mass function.} interacts with a much smaller secondary. If the mass ratio is very large, quasar activity is not supposed to be triggered at the level of observability our sample requires. If the mass ratio is $\simgt$ 4, quasar activity is expected \citep{Colpi2007}, but the delay between the onset in the primary and in the secondary is too long for the pair to be detected as a ``double quasar". Hence we prefer the third hypothesis above.

Clearly the occupation fraction (that is the fraction of galaxies hosting a MBH) is indeed an important factor. If not all galaxies host a MBH the fraction of binaries decreases. As long as we do not have a demography of inactive MBHs at $z>0$, this issue cannot be addressed properly.  This fair comparison has to await for the Laser Interferometric Space Antenna, probing the BH population via gravitational waves detections.  However, we can compare our hypothesis to theoretical expectations. In quasar evolution models that evolve the population {\it ab initio} a minimum number density of MBHs is required to match the luminosity function of quasars. That is, one can not have less MBHs than quasars, at a given time. With this constraint in mind  models such as those described in Volonteri et al. 2008 (see also Marulli et al. 2007) require an occupation fraction higher than 90\% for MBHs hosted in halos with mass above  $10^{12}\msun$, the typical quasar host halo found in clustering analysis \citep{MyersB,Coil}.

We compared the typical separations at which binary quasars are observed  (Figure~\ref{fig:sep}) to the merger phase and separations that should correspond to the quasar phase in merger simulations. We find that the distribution of physical separations peaks at $\sim 30$ kpc, with a tail at larger separations ($\sim 100-200$ kpc). The peak of the distribution is roughly consistent with the first episode of quasar activity found in equal mass mergers simulations  \citep[see, Figure 2 in][]{HopkinsC}.  The tail of the quasar pairs distribution at large separations does not correspond to any quasar phase predicted by published simulation results. However, the mass ratio of the black holes in the pairs does not match the 1:1 ratio that was typically selected in galaxy merger simulations. The average mass ratio is $\langle 2.4\pm1.3\rangle$, a regrettably still unexplored range. 

The pairs at large separation could be unequal mass mergers  seen at the first apocentric passage.  This scenario would follow naturally if gas is dynamically perturbed during the first pericentric passage, but the gas in the disk requires longer timescales to loose enough angular momentum to be driven at subparsec scales and fuel the MBH. This final stage likely occurs when the two galaxies are at the first apocenter. 

Finally, we explored how the redhift dependence (or independence) of the correlations between MBHs and their hosts can be tested using large samples of quasar pairs. If MBHs follow a redshift independent correlation with either the velocity dispersion of the host (e.g., Tremaine et al. 2002, Ferrarese \& Merritt 2000, case A), or with the mass of the bulge (e.g. Marconi \& Hunt 2004, Haring \& Rix 2004, case B), we expect merger timescales to decrease at high redshift, leading to a higher fraction of double quasars. If, on the other hand, the correlations between MBHs and their hosts evolve with time (cases C and D, e.g., McLure et al. 2006, Woo et al. 2006) the merger timescales vary extremely slowly with cosmic time, and an increase in the binary fraction is not expected.


\acknowledgements
We wish to thank Marianne Vestergaard for kindly providing black hole masses for SDSS Data Release 3 quasars.


\newpage

\LongTables
\begin{deluxetable}{lcccccccc}
\tabletypesize{\footnotesize}
\tablenum{1}
\tablecaption{Properties of the double quasar sample. We list the name and redshift of the pair, the physical separation between the two quasars, along with the luminosity and BH mass for both quasars in the pair. The last column lists the reference for the photometric or spectroscopic information. (1) Hennawi et al. (2006), (2) Myers et al. (2007a), (3) Myers et al. (2007b), (A) Vestergaard et al. (2008)}
\tablehead{\colhead{Name} & \colhead{z}  &  \colhead{R$_{max}$} & \colhead{L$_B$ } & \colhead{L$_B$} & \colhead{BH mass} & \colhead{BH mass} & \colhead{Ref.*} \\ 
\colhead{} & \colhead{} & \colhead{} & \colhead{quasar 1} & \colhead{quasar 2} & \colhead{quasar 1} & \colhead{quasar 2} & \colhead{} \\
\colhead{} & \colhead{} &  \colhead{(kpc)} & \colhead{($10^{12}$L$_\odot$)} & \colhead{($10^{12}$L$_\odot$)} & \colhead{($10^9$M$_\odot$)} & \colhead{($10^9$M$_\odot$)} & \colhead{} } 
\startdata
SDSSJ1120+6711 & 1.49 & 12.69 & 3.09$\pm{1.72}$ & 7.72$\pm{0.05}$E-1 & 7.32$\pm1.16$E-1 & 2.23$\pm{0.29}$E-1 & 1, A\\
SDSSJ2336-0107 & 1.29 & 14.23 & 9.37     $\pm{0.61}$E-1 & 3.90$\pm{0.02}$E-1 & 5.67$\pm^{10.81}_{3.86}$E-1 & 1.18$\pm^{4.50}_{1.61}$E-1 & 1\\
SDSSJ0048-1051 & 1.56 & 30.49 & 1.06     $\pm{0.06}$ & 1.68   $\pm{0.09}$ & 3.57$\pm^{2.11}_{1.33}$E-1 & 9.93$\pm^{18.92}_{6.77}$E-1 & 1, A\\
SDSSJ0054-0946 & 2.13  & 117.11 & 2.63$\pm{0.15}$E+1 & 2.02   $\pm{0.11}$ & 2.24$\pm^{0.25}_{0.22}$ & 1.19$\pm^{2.26}_{0.81}$ & 1, A\\
SDSSJ0117+3153 & 2.13  & 93.85 & 2.38    $\pm{0.13}$ & 4.43   $\pm{0.25}$ & 1.39$\pm^{2.65}_{0.95}$ & 2.52$\pm^{4.82}_{1.73}$ & 1\\
SDSSJ0201+0032 & 2.30  & 155.90 & 7.63   $\pm{0.42}$ & 3.35   $\pm{0.19}$ & 4.28$\pm^{8.17}_{2.93}$ & 1.93$\pm^{3.68}_{1.32}$ & 1\\
SDSSJ0248+0009 & 1.64  & 58.45 & 4.07    $\pm{0.23}$ & 1.07   $\pm{0.06}$ & 1.68$\pm^{0.85}_{0.56}$ & 6.47$\pm^{12.32}_{4.40}$E-1 & 1, A\\
SDSSJ0332-0722 & 2.10  & 150.62 & 2.95   $\pm{0.16}$ & 2.24   $\pm{0.12}$ & 1.71$\pm^{3.26}_{1.17}$ & 1.31$\pm^{2.50}_{0.89}$ & 1\\
SDSSJ0846+2749 & 2.12  & 156.25 & 5.84   $\pm{0.32}$ & 4.41   $\pm{0.24}$ & 3.30$\pm^{6.30}_{2.26}$ & 2.52$\pm^{4.81}_{1.72}$ & 1\\
SDSSJ0939+5953 & 2.53  & 262.47 & 6.34   $\pm{0.35}$ & 1.85$\pm{0.10}$E+1 & 1.15$\pm 0.40$ & 1.02$\pm^{1.94}_{0.70}$E+1 & 1, A\\
SDSSJ0955+6045 & 0.72  & 134.40 & 1.81$\pm{0.10}$E-1 & 1.50$\pm{0.08}$E-1 & 8.01$\pm 4.26$E-1 & 9.98$\pm^{19.00}_{6.74}$E-2 & 1, A\\
SDSSJ0959+5449 & 1.95  & 32.73 & 2.93    $\pm{0.16}$ & 1.86   $\pm{0.10}$ & 1.69$\pm^{3.23}_{1.16}$ & 1.09$\pm^{2.08}_{0.75}$ & 1\\
SDSSJ1010+0416 & 1.51  & 145.59 & 1.56   $\pm{0.09}$ & 1.41   $\pm{0.08}$ & 3.40$\pm 2.16$E-1 & 8.37$\pm^{15.95}_{5.70}$E-1 & 1, A\\
SDSSJ1014+0920 & 2.29  & 180.65 & 9.19   $\pm{0.51}$ & 3.39   $\pm{0.19}$ & 1.17$\pm 0.26$ & 1.95$\pm^{3.72}_{1.33}$ & 1, A\\
SDSSJ1028+3929 & 1.89  & 63.12 & 3.21    $\pm{0.18}$ & 1.38   $\pm{0.08}$ & 1.85$\pm^{3.53}_{1.27}$ & 8.21$\pm^{15.64}_{5.59}$E-1 & 1\\
SDSSJ1034+0701 & 1.25  & 25.85 & 1.63    $\pm{0.09}$ & 3.37$\pm{0.19}$E-1 & 1.43$\pm 0.22$ & 2.15$\pm^{4.09}_{1.45}$E-1 & 1, A\\
SDSSJ1056-0059 & 2.13  & 59.80 & 4.11    $\pm{0.23}$ & 1.94   $\pm{0.11}$ & 2.35$\pm^{4.48}_{1.61}$ & 1.14$\pm^{2.17}_{0.78}$ & 1\\
SDSSJ1123+0037 & 1.17  & 465.12 & 2.22   $\pm{0.12}$ & 6.96$\pm{0.39}$E-1 & 4.83$\pm 1.09$E-1 & 4.27$\pm^{8.14}_{2.90}$E-1 & 1, A\\
SDSSJ1225+5644 & 2.38 & 48.92 & 8.93     $\pm{0.49}$ & 2.95   $\pm{0.16}$ & 4.99$\pm^{9.52}_{3.42}$ & 1.71$\pm^{3.25}_{1.16}$ & 1\\
SDSSJ1254+6104 & 2.05 & 146.91 & 8.33    $\pm{0.46}$ & 5.32   $\pm{0.29}$ & 3.60$\pm 1.31$ & 3.02$\pm^{5.76}_{2.06}$ & 1, A\\
SDSSJ1259+1241 & 2.19 & 29.78 & 4.06     $\pm{0.22}$ & 4.42   $\pm{0.24}$ & 2.32$\pm^{4.43}_{1.59}$ & 2.52$\pm^{4.81}_{1.73}$ & 1\\
SDSSJ1303+5100 & 1.68 & 32.18 & 1.60     $\pm{0.09}$ & 1.22   $\pm{0.07}$ & 9.47$\pm^{18.06}_{6.46}$E-1 & 7.31$\pm^{13.92}_{4.97}$E-1 & 1\\
SDSSJ1310+6208 & 2.06 & 391.25 & 1.12 $\pm{0.06}$E+1 & 2.27   $\pm{0.13}$ & 2.52$\pm 0.55$ & 1.32$\pm^{2.52}_{0.90}$ & 1, A\\
SDSSJ1337+6012 & 1.73 & 26.23 & 8.73     $\pm{0.48}$ & 2.31   $\pm{0.13}$ & 2.95$\pm^{1.09}_{0.80}$ & 1.35$\pm^{2.57}_{0.92}$ & 1, A\\
SDSSJ1349+1227 & 1.72 & 25.39 & 1.80  $\pm{0.10}$E+1 & 4.46   $\pm{0.25}$ & 9.87$\pm^{18.83}_{6.77}$ &2.55$\pm^{4.86}_{1.74}$ & 1\\
SDSSJ1400+1232 & 2.05 & 121.87 & 2.73    $\pm{0.15}$ & 2.40   $\pm{0.13}$ & 1.59$\pm^{3.02}_{1.08}$ & 1.40$\pm^{2.67}_{0.96}$ & 1\\
SDSSJ1405+4447 & 2.23  & 61.04 & 2.26 $\pm{0.12}$E+1 & 3.67   $\pm{0.20}$ & 1.23$\pm^{2.35}_{0.84}$E+1 & 2.11$\pm^{4.02}_{1.44}$ & 1\\
SDSSJ1409+3919 & 2.08  & 56.66 & 2.96    $\pm{0.16}$ & 1.61   $\pm{0.09}$ & 1.72$\pm^{3.27}_{1.17}$ & 9.53$\pm^{18.15}_{6.49}$E-1 & 1\\
SDSSJ1530+5304 & 1.53  & 34.72 & 1.02    $\pm{0.06}$ & 9.55$\pm{0.53}$E-1 & 6.13$\pm^{11.68}_{4.17}$E-1 & 5.78$\pm^{11.01}_{3.93}$E-1 & 1\\
SDSSJ1546+5134 & 2.95  & 326.69 & 3.91   $\pm{0.22}$ & 1.22$\pm{0.07}$E+1 & 2.24$\pm^{4.28}_{1.53}$ & 6.77$\pm^{12.90}_{4.63}$ & 1\\
SDSSJ1629+3724 & 0.92  & 34.49 & 9.56 $\pm{0.53}$E-1 & 8.48$\pm{0.47}$E-1 & 6.46$\pm 2.53$E-1 & 5.16$\pm^{9.83}_{3.51}$E-1 & 1, A\\
SDSSJ1719+2549 & 2.17  & 121.76 & 4.26   $\pm{0.24}$ & 4.99   $\pm{0.28}$ & 1.17$\pm 0.77$ & 2.84$\pm^{5.42}_{1.94}$ & 1, A\\
SDSSJ1723+5904 & 1.60  & 31.35 & 4.49    $\pm{0.25}$ & 1.16   $\pm{0.06}$ & 2.28$\pm^{0.83}_{0.61}$ & 6.98$\pm^{13.30}_{4.75}$E-1 & 1, A\\
SDSSJ2128-0617 & 2.07  & 69.20 & 4.91    $\pm{0.27}$ & 4.48   $\pm{0.25}$ & 2.80$\pm^{5.33}_{1.91}$ & 2.56$\pm^{4.88}_{1.75}$ & 1\\
SDSSJ2214+1326 & 2.00  & 48.55 & 2.76    $\pm{0.15}$ & 2.24   $\pm{0.12}$ & 1.60$\pm^{3.06}_{1.09}$ & 1.31$\pm^{2.50}_{0.89}$ & 1\\
SDSSJ2220+1247 & 1.99  & 133.17 & 3.22   $\pm{0.18}$ & 1.52   $\pm{0.08}$ & 1.86$\pm^{3.54}_{1.27}$ & 9.00$\pm^{17.15}_{6.13}$E-1 & 1\\
SDSSJ0740+2926 & 0.98  & 20.72 & 2.36    $\pm{0.13}$ & 7.54$\pm{0.42}$E-1 & 4.14$\pm 0.62$E-1 & 4.61$\pm^{8.78}_{3.13}$E-1 & 1, A\\
SDSSJ1035+0752 & 1.22  & 22.45 & 2.22    $\pm{0.12}$ & 6.61$\pm{0.37}$E-1 & 5.79$\pm 1.93$E-1 & 4.07$\pm^{7.75}_{2.76}$E-1 & 1, A\\
SDSSJ1124+5710 & 2.31  & 18.04 & 1.49 $\pm{0.08}$E+1 & 5.24   $\pm{0.29}$ & 2.55$\pm 0.34$ & 2.97$\pm^{5.67}_{2.03}$ & 1, A\\
SDSSJ1138+6807 & 0.77  & 19.26 & 1.98    $\pm{0.11}$ & 3.75$\pm{0.21}$E-1 & 1.76$\pm^{0.21}_{0.19}$E-1 & 2.37$\pm^{4.51}_{1.61}$E-1 & 1, A\\
SDSSJ1508+3328 & 0.88  & 22.45 & 3.41    $\pm{0.19}$ & 3.11$\pm{0.17}$E-1 & 1.96$\pm^{3.74}_{1.34}$ & 1.99$\pm^{3.79}_{1.35}$E-1 & 1\\
SDSSJ0012+0052 & 1.63  & 135.55 & 8.64$\pm{0.48}$E-1 & 9.34$\pm{0.52}$E-1 & 5.25$\pm^{10.00}_{3.57}$E-1 & 5.66$\pm^{10.78}_{3.85}$E-1 & 1\\
SDSSJ0117+0020 & 0.61  & 299.78 & 1.49   $\pm{0.08}$ & 1.74$\pm{0.10}$E-1 & 2.26$\pm^{0.22}_{0.20}$E-1 & 1.15$\pm^{2.19}_{0.78}$E-1 & 1, A\\
SDSSJ0141+0031 & 1.89  & 361.05 & 2.47   $\pm{0.14}$ & 1.67   $\pm{0.09}$ & 1.44$\pm^{2.74}_{0.98}$ & 9.88$\pm^{18.83}_{6.73}$E-1 & 1\\
SDSSJ0245-0113 & 2.46  & 36.45 & 5.78    $\pm{0.32}$ & 3.04   $\pm{0.17}$ & 3.27$\pm^{6.24}_{2.24}$ & 1.76$\pm^{3.35}_{1.20}$ & 1\\
SDSSJ0258-0003 & 1.32  & 246.69 & 3.27   $\pm{0.18}$ & 1.40   $\pm{0.08}$ & 8.36$\pm 2.13$E-1 & 8.32$\pm^{15.86}_{5.67}$E-1 & 1, A\\
SDSSJ0259+0048 & 0.89  & 152.21 & 8.59$\pm{0.48}$E-1 & 1.07   $\pm{0.06}$ & 1.60$\pm 0.94$ & 6.42$\pm^{12.23}_{4.37}$E-1 & 1, A\\
SDSSJ0350-0031 & 2.00  & 380.88 & 2.82   $\pm{0.16}$ & 5.66   $\pm{0.31}$ & 1.64$\pm^{3.12}_{1.12}$ & 3.21$\pm^{6.12}_{2.19}$ & 1\\
SDSSJ0743+2054 & 1.56  & 300.71 & 2.03   $\pm{0.11}$ & 1.64   $\pm{0.09}$ & 3.63$\pm 1.52$E-1 & 9.70$\pm^{18.48}_{6.61}$E-1 & 1, A\\
SDSSJ0747+4318 & 0.50  & 56.16 & 2.17 $\pm{0.12}$E-1 & 1.78$\pm{0.10}$E-1 & 2.85$\pm^{0.94}_{0.70}$E-1 & 1.17$\pm^{2.23}_{0.79}$E-1 & 1, A\\
SDSSJ0824+2357 & 0.54  & 94.66 & 5.18 $\pm{0.29}$E-1 & 4.39$\pm{0.24}$E-1 & 3.23$\pm^{6.14}_{2.19}$E-1 & 2.75$\pm^{5.25}_{1.87}$E-1 & 1\\
SDSSJ0856+5111 & 0.54  & 138.49 & 5.00$\pm{0.28}$E-1 & 1.91$\pm{0.11}$E-1 & 1.79$\pm^{0.22}_{0.20}$E-1 & 1.43$\pm^{0.54}_{0.40}$ & 1, A\\
SDSSJ0909+0002 & 1.87  & 126.35 & 6.16$\pm{0.34}$E+1 & 2.72   $\pm{0.15}$ & 5.92$\pm^{0.95}_{0.82}$ & 9.41$\pm^{15.05}_{9.41}$E-1 & 1\\
SDSSJ0955+0616 & 1.28  & 368.00 & 6.99   $\pm{0.39}$ & 8.58$\pm{0.48}$E-1 & 2.08$\pm 0.39$ & 5.21$\pm^{9.93}_{3.54}$E-1 & 1, A\\
SDSSJ1032+0140 & 1.46  & 465.77 & 4.42   $\pm{0.25}$ & 1.02   $\pm{0.06}$ & 7.14$\pm 1.54$E-1 & 6.14$\pm^{11.70}_{4.18}$E-1 & 1, A\\
SDSSJ1103+0318 & 1.94  & 481.14 & 1.43$\pm{0.08}$E+1 & 7.58   $\pm{0.42}$ & 2.27$\pm^{0.81}_{0.60}$ & 4.88$\pm 2.23$ & 1, A\\
SDSSJ1107+0033 & 1.88  & 208.81 & 7.70   $\pm{0.43}$ & 2.83   $\pm{0.16}$ & 4.32$\pm^{8.24}_{2.96}$ & 1.64$\pm^{3.12}_{1.12}$ & 1\\
SDSSJ1116+4118 & 2.99  & 106.41 & 2.63$\pm{0.15}$E+1 & 1.12$\pm{0.06}$E+1 & 1.43$\pm^{2.72}_{0.98}$E+1 & 6.24$\pm^{11.90}_{4.27}$ & 1\\
SDSSJ1134+0849 & 1.53  & 229.47 & 3.90   $\pm{0.22}$ & 2.72   $\pm{0.15}$ & 1.19$\pm 0.36$ & 1.58$\pm^{3.01}_{1.08}$ & 1, A\\
SDSSJ1146-0124 & 1.98  & 238.83 & 2.55   $\pm{0.14}$ & 2.15   $\pm{0.12}$ & 1.49$\pm^{2.82}_{1.01}$ & 1.26$\pm^{2.39}_{0.86}$ & 1\\
SDSSJ1152-0030 & 0.55  & 187.86 & 4.19$\pm{0.23}$E-1 & 1.23$\pm{0.07}$E-1 & 5.01$\pm 0.88$E-1 & 8.33$\pm^{15.85}_{5.62}$E-2 & 1, A\\
SDSSJ1207+0115 & 0.97  & 281.41 & 1.49   $\pm{0.08}$ & 3.81$\pm{0.21}$E-1 & 2.95$\pm 0.57$E-1 & 2.41$\pm^{4.58}_{1.63}$E-1 & 1, A\\
2QZJ1217+0006 & 1.78  & 437.70 & 2.71    $\pm{0.15}$ & 4.96   $\pm{0.27}$ & 1.57$\pm^{3.00}_{1.07}$ & 2.82$\pm^{5.38}_{1.93}$ & 1\\
2QZJ1217+0055 & 0.90  & 315.54 & 3.74 $\pm{0.21}$E-1 & 4.84$\pm{0.27}$E-1 & 2.37$\pm^{4.51}_{1.61}$E-1 & 3.02$\pm^{5.75}_{2.05}$E-1 & 1\\
SDSSJ1226-0112 & 0.920 & 396.67 & 5.46   $\pm{0.30}$ & 5.97$\pm{0.33}$E-1 & 9.75$\pm 0.18$E-1 & 3.69$\pm^{7.03}_{2.51}$E-1 & 1, A\\
SDSSJ1300-0156 & 1.620 & 377.02 & 1.00$\pm{0.06}$E+1 & 2.58   $\pm{0.14}$ & 7.84$\pm^{0.95}_{0.85}$E-1 & 1.50$\pm^{2.86}_{1.02}$ & 1, A\\
2QZJ1328-0157 & 2.37  & 429.23 & 7.76    $\pm{0.43}$ & 5.08   $\pm{0.28}$ & 4.35$\pm^{8.30}_{2.98}$ & 2.89$\pm^{5.51}_{1.98}$ & 1\\
2QZJ1354-0108 & 1.99  & 464.84 & 5.17    $\pm{0.29}$ & 2.85   $\pm{0.16}$ & 2.94$\pm^{5.60}_{2.01}$ & 1.65$\pm^{3.15}_{1.13}$ & 1\\
SDSSJ0846+2710 & 2.17  & 38.42 & 2.15    $\pm{0.12}$ & 2.23   $\pm{0.12}$ & 1.26$\pm^{2.39}_{0.86}$ & 1.30$\pm^{2.48}_{0.89}$ & 2\\
SDSSJ0927+0707 & 1.38  & 40.59 & 9.11 $\pm{0.50}$E-1 & 5.70$\pm{0.31}$E-1 & 5.52$\pm^{10.52}_{3.76}$E-1 & 3.53$\pm^{6.72}_{2.40}$E-1 & 2\\
SDSSJ1004+4112 & 1.73  & 31.82 & 8.09    $\pm{0.45}$ & 5.53   $\pm{0.31}$ & 4.53$\pm^{8.64}_{3.10}$ & 1.59$\pm^{0.49}_{0.37}$ & 2, A\\
SDSSJ1007+0126 & 1.23  & 34.85 & 3.80 $\pm{0.21}$E-1 & 6.03$\pm{0.33}$E-1 & 2.40$\pm^{4.58}_{1.63}$E-1 & 3.73$\pm^{7.10}_{2.53}$E-1 & 2\\
SDSSJ1207+1408 & 1.77  & 33.38 & 1.71    $\pm{0.09}$ & 1.91   $\pm{0.11}$ & 1.01$\pm^{1.92}_{0.69}$ & 1.13$\pm^{2.15}_{0.77}$ & 2\\
SDSSJ1431+2705 & 2.38  & 48.21 & 3.73    $\pm{0.21}$ & 5.56   $\pm{0.31}$ & 2.14$\pm^{4.08}_{1.46}$ & 3.15$\pm^{6.01}_{2.16}$ & 2\\
SDSSJ1447+6327 & 1.23  & 33.19 & 4.02 $\pm{0.22}$E-1 & 1.57   $\pm{0.09}$ & 2.53$\pm^{4.83}_{1.72}$E-1 & 9.32$\pm^{17.77}_{6.35}$E-1 & 2\\
SDSSJ1649+1733 & 2.08  & 30.17 & 6.59    $\pm{0.36}$ & 4.43   $\pm{0.24}$ & 3.72$\pm^{7.09}_{2.54}$ & 2.53$\pm^{4.82}_{1.73}$ & 2\\
SDSSJ1158+1235 & 0.60  & 23.73 & 1.77 $\pm{0.10}$E-1 & 1.43$\pm{0.08}$ E-1& 1.17$\pm^{2.23}_{0.79}$E-1 & 9.54$\pm^{18.17}_{6.45}$E-2 & 3\\
SDSSJ1320+3056 & 1.60  & 40.16 & 6.83    $\pm{0.38}$ & 1.97   $\pm{0.11}$ & 3.85$\pm^{7.34}_{2.63}$ & 1.16$\pm^{2.20}_{0.79}$ & 3\\
SDSSJ1418+2441 & 0.57  & 29.44 & 2.82 $\pm{0.16}$E-1 & 1.16$\pm{0.06}$E-1 & 1.81$\pm^{3.45}_{1.23}$E-1 & 7.88$\pm^{15.00}_{5.32}$E-2 & 3\\
SDSSJ1426+0719 & 1.31  & 35.80 & 1.08    $\pm{0.06}$ & 5.68$\pm{0.31}$E-1 & 6.49$\pm^{12.37}_{4.42}$E-1 & 3.52$\pm^{6.71}_{2.39}$E-1 & 3\\
SDSSJ1430+0714 & 1.25  & 45.14 & 1.69    $\pm{0.09}$ & 7.98$\pm{0.44}$E-1 & 9.97$\pm^{19.00}_{6.80}$E-1 & 4.87$\pm^{9.28}_{3.31}$E-1 & 3\\
SDSSJ1458+5448 & 1.91  & 43.21 & 1.79    $\pm{0.10}$ & 1.41   $\pm{0.08}$ & 1.05$\pm^{2.01}_{0.72}$ & 8.37$\pm^{15.94}_{5.70}$E-1 & 3\\
SDSSJ1606+2900 & 0.77  & 25.55 & 1.33    $\pm{0.07}$ & 1.21   $\pm{0.07}$ & 7.91$\pm^{15.07}_{5.39}$E-1 & 7.23$\pm^{13.77}_{4.92}$E-1 & 3\\
SDSSJ1635+2911 & 1.59  & 41.68 & 1.23    $\pm{0.07}$ & 5.37   $\pm{0.30}$ & 7.36$\pm^{14.03}_{5.01}$E-1 & 3.05$\pm^{5.82}_{2.09}$ & 3\\
\enddata
\end{deluxetable}

\end{document}